\documentclass{mem}
\pdfoutput=1
\usepackage{natbib}
\usepackage{txfonts}
\usepackage{balance}
\usepackage{graphicx}
\usepackage{hyperref}
\usepackage{epstopdf}
\idline{0}{1}

\newcommand{\Ms}{\ensuremath{M_{\odot}}}

\newcommand{\cf}{{\it c.f.~}}
\newcommand{\ie}{{\it i.e.}}

\newcommand{\beq}{\begin{equation}}
\newcommand{\eeq}{\end{equation}}

\newcommand{\mcl}{\ensuremath{M_{cl}}}
\newcommand{\rh}{\ensuremath{r_h}}

\begin{document}

\title{
Blue straggler formation at core collapse 
}

   \subtitle{}

\author{
Sambaran Banerjee\inst{1} 
          }

\institute{
AIfA/HISKP, University of Bonn,
Auf dem H\"ugel 71,
D-53121 Bonn,
Germany
}

\authorrunning{Banerjee}

\titlerunning{BSS formation at core collapse}

\abstract{
Among the most striking
feature of blue straggler stars (BSS)
in globular clusters is the presence of multiple sequences of BSSs in the
colour-magnitude diagrams (CMDs) of several globular clusters. It is often
envisaged that such a multiple BSS sequence would arise due a recent core collapse
of the host cluster, triggering a number of stellar collisions and
binary mass transfers simultaneously over a brief episode of time.
Here we examine this scenario using direct N-body computations of
moderately-massive star clusters (of order $10^4\Ms$). As a preliminary attempt,
these models are initiated with $\approx8-10$ Gyr old stellar population and
King profiles of high concentrations, being ``tuned'' to undergo core collapse quickly.
BSSs are indeed found to form in a ``burst'' at the onset of the core collapse
and several of such BS-bursts occur during the post-core-collapse phase.
In those models that include a few percent primordial binaries,
both collisional and binary BSSs form after the onset of the (near) core-collapse.
However, there is as such no clear discrimination between the two types of BSSs in
the corresponding computed CMDs. We note that this may be due to the
less number of BSSs formed in these less massive models than that in actual globular clusters.
}
\maketitle{}

\section{Introduction}\label{intro}

Blue Straggler Stars (hereafter BSS) of a stellar ensemble
are main-sequence (hereafter MS) stars that are located on the colour-magnitude-diagram
(hereafter CMD) as an extension of the regular MS, beyond
its turn-off point. The most natural explanation for this stellar
subpopulation is that they are stars with rejuvenated hydrogen
content so that they are still on the MS despite being
more massive than the turnoff mass. There are two primary
channels considered for this rejuvenation and mass gain, namely, (a)
direct stellar collisions \citep{hd76} and
(b) mass transfer in a binary \citep{hur2001,hur2005}.    

Being distinctly identifiable in the CMD and being more massive than
the average stellar population, BSSs serve as excellent
tracers of dynamical processes in open and globular clusters
\citep{mg2009,fr2012}; especially their radial profile, as
governed by their mass segregation, serves
as a ``dynamical clock'' for globular clusters \citep{fr2012}. 

The focus of this study is another remarkable feature of the BSSs,
namely, the existence of a double sequence of BSSs in several
globular clusters. Perhaps the best example of this is
the BSSs' double sequence observed in the globular cluster
M30 \citep{fr2009}; another vivid example is that of
the globular NGC 362 \citep{dal2013}. An intriguing explanation
for this is the recent core collapse of the parent cluster
that triggers both of the formation channels simultaneously
so that the ``red'' sequence comprise binary (mass-transferring)
BSSs and the ``blue'' sequence comprise the collision products
\citep{fr2009}. Indeed, the presence of a central cusp
in the density profile of M30 does indicate its post core collapse
state \citep{fr2009}. 
More recent and detailed study of mass transfer
in binary evolution models only strengthens this notion and
constrain the types of binaries that make up the red
sequence \citep{xin2015}.
Generally, the red BSSs are found to be more
centrally concentrated than the blue ones. 

Motivated by the above possibility, the objective of the present
work is to study the BSS formation following core collapse
\citep{spitzer1987,hh2003} of a model star cluster. To allow the stars
to collide ``naturally'' (\ie, without any constructed collision-triggering
procedure), a direct N-body approach is followed. This is
for the first time stellar collisions and BSS formation
following core collapse (or, speaking more generally,
during the central energy-generation phase of a cluster; \citealt{spitzer1987})
is studied explicitly.

\begin{figure}
\includegraphics[width=6.5cm, angle=0, clip=true]{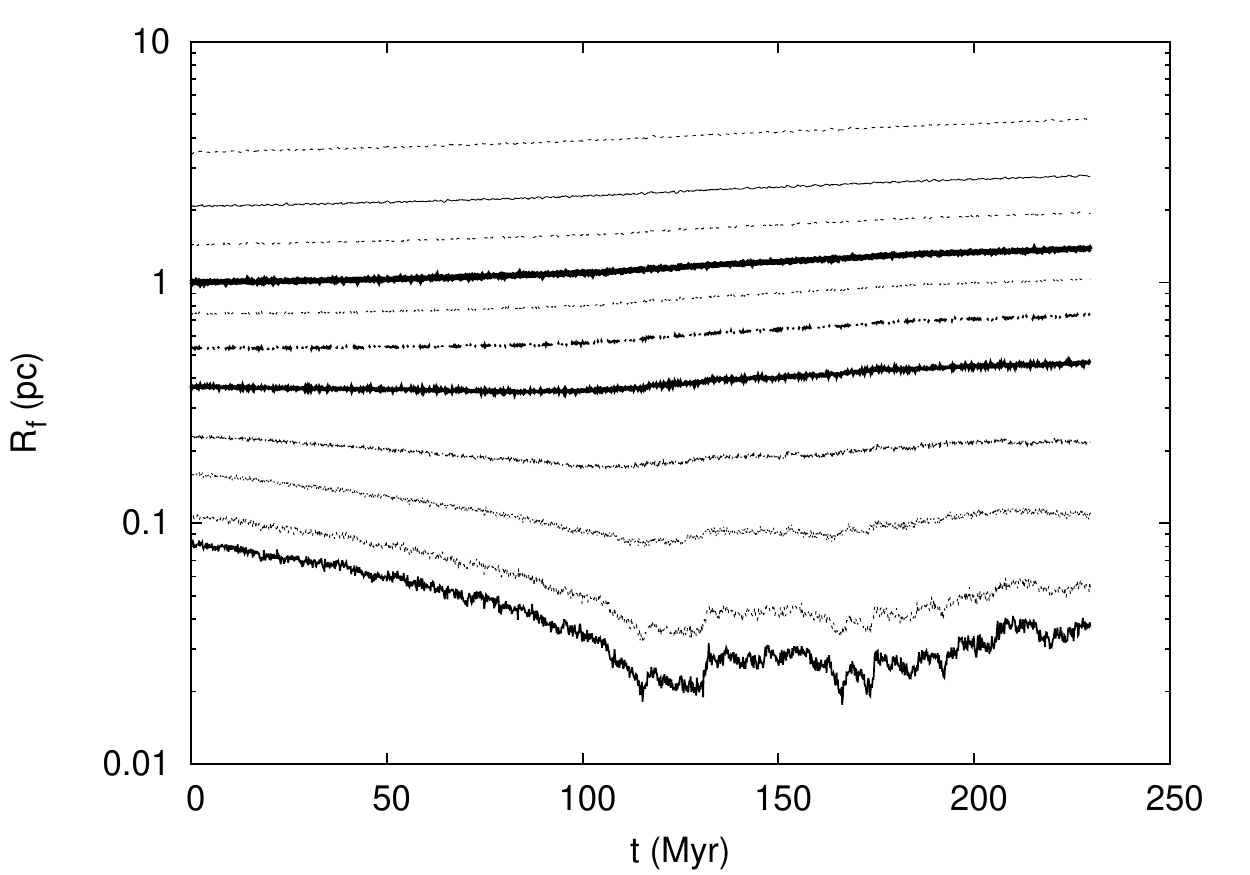}
\includegraphics[width=6.5cm, angle=0, clip=true]{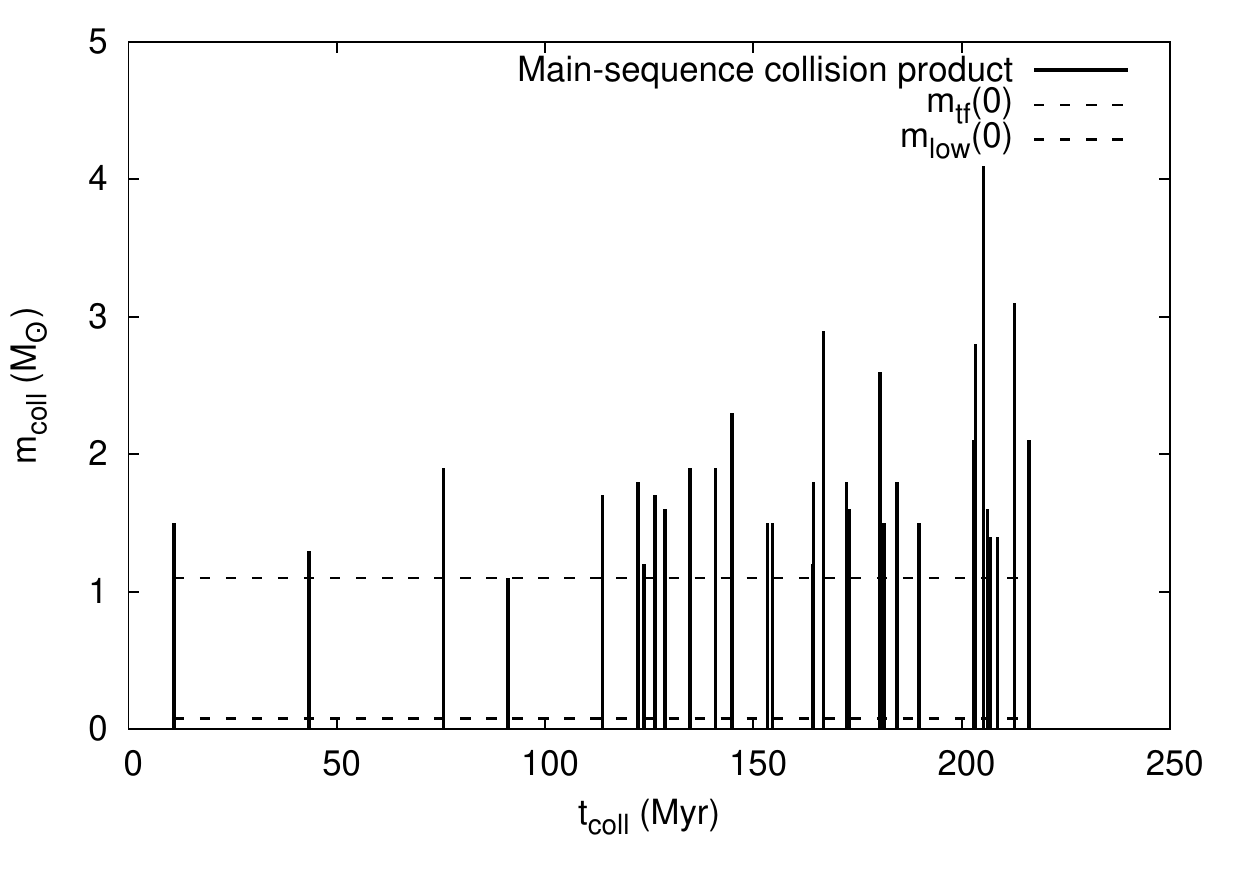}
\vspace{-0.9 cm}
\caption{\footnotesize The Lagrangian radii (top) and the time sequence of formation
of MS collision products (bottom) for the computation with
$\mcl(0)\approx1.5\times10^4\Ms$, $W_0=7.5$ and $\rh(0)=1$ pc
with initially only single stars.}
\label{fig:15K_single_evol}
\end{figure}

\section{N-body calculations}\label{calc} 
\vspace{-0.32 cm}

To obtain a prompt but clear core collapse, systems with a moderate
number, $N$, of stars and having high central concentration needs
to be evolved --- that way one ``tunes'' a cluster for core collapse.
In this work, clusters initially with masses, $\mcl(0)$, between $1-3\times10^4\Ms$
following $W_0=7.5$ \citet{king1962} density profiles and of half-mass
radius $\rh(0)\approx1$ pc are computed. Such concentrations are common
in present-day Galactic and Local-Group globulars. To mimic an old stellar
population like in globular clusters, low-mass
stars between $0.1-1.0\Ms$ with a \citet{krp2001} mass function is
assumed, which are pre-evolved for $\approx8.8$ Gyr (this
corresponds to the ``turn-off'' age of $1.0\Ms$ stars at
solar metallicity, when the
giant branch on the CMD is just appearing). Such models are computed
both with initially single stars and with $\approx5$\% binaries
following a \citet{dq91} period distribution. No initial mass
segregation is assumed as consistent with what is typically
observed in globulars. All the N-body runs
are done using the {\tt NBODY7} code \citep{aseth2012,nita2012}.
\vspace{-0.57 cm}

\section{Results}\label{res}

Fig.~\ref{fig:15K_single_evol} (top panel) shows the evolution of the
Lagrangian radii of such a computed model
with $\mcl(0)\approx1.5\times10^4\Ms$ without initial binaries.
The occurrence of core collapse is indicated by the
abrupt halting of the inner region of the cluster, followed
by a slow expansion. The high central density
at and after the collapse boosts collisions among stars
which can be called a ``burst'';
Fig.~\ref{fig:15K_single_evol} (bottom) shows the timeline for collisions occurring
among MS stars where BSSs are those whose final
mass (Y-axis) exceeds the turn-off mass (upper horizontal line). 

\begin{figure}
\flushleft
\includegraphics[width=5.2cm, angle=-90, clip=true]{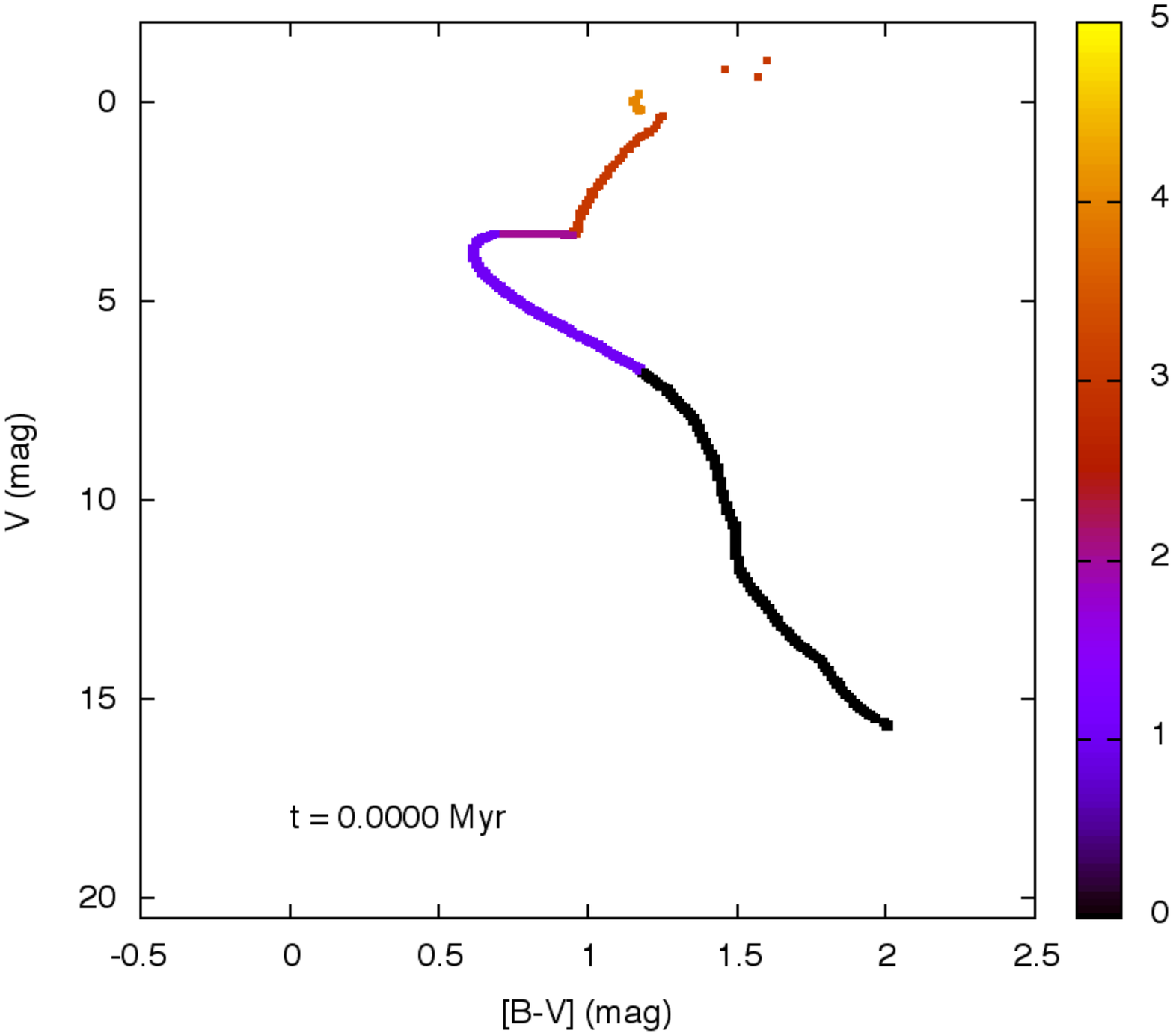}
\includegraphics[width=5.2cm, angle=-90, clip=true]{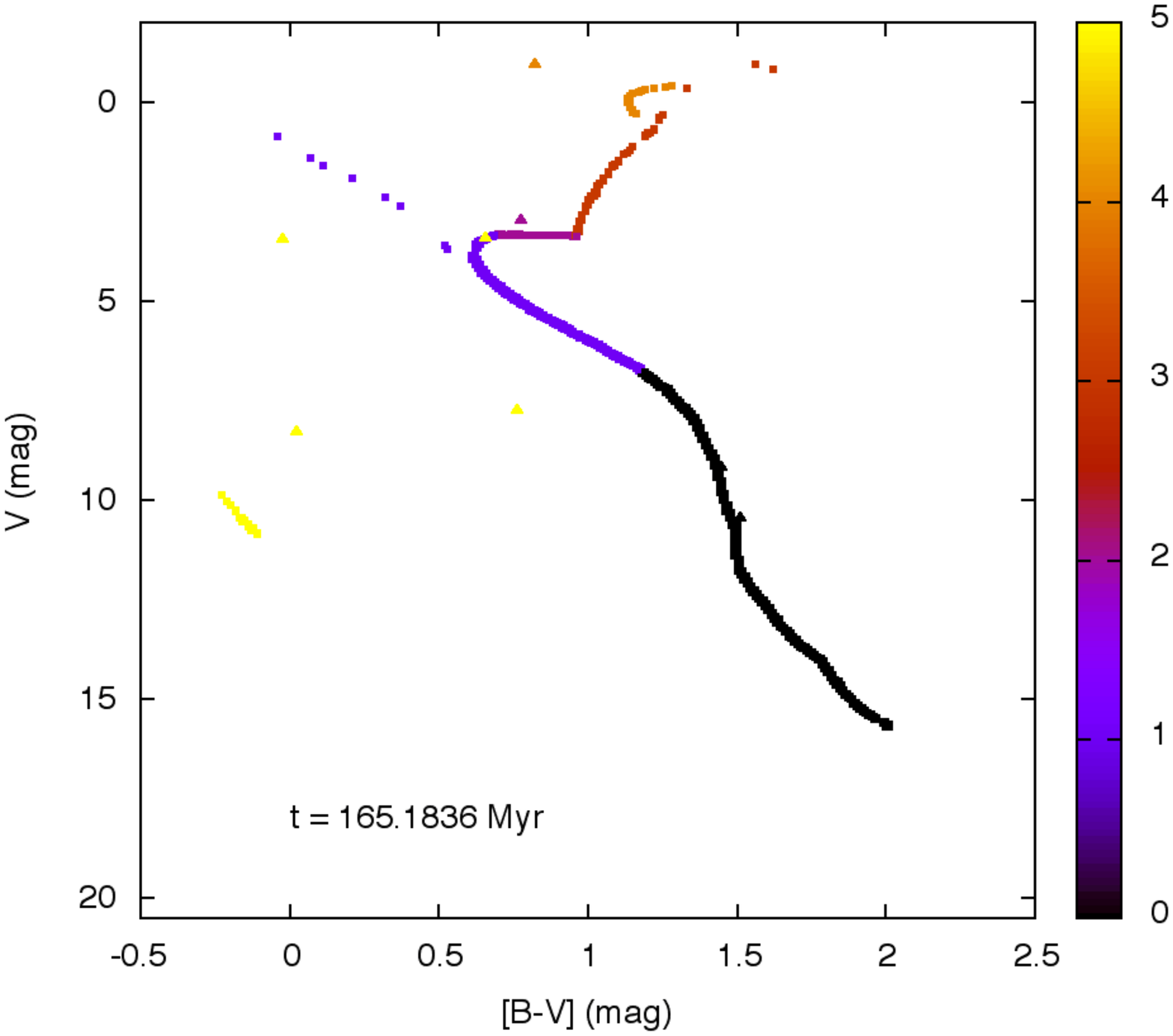}
\vspace{-0.4 cm}
\caption{\footnotesize CMDs corresponding to the calculation in
Fig.~\ref{fig:15K_single_evol}. The colour coding
represents the different stellar-evolutionary stages where
0=lower MS, 1=upper MS, 2=subgiant, 3=red giant, etc. A filled square
indicates a single star and a filled triangle implies a binary (combined magnitudes
used).}
\label{fig:15K_single_cmd}
\end{figure}

Fig.~\ref{fig:15K_single_cmd} shows the corresponding computed CMDs at $t=0$ Myr and
$t\approx165$ Myr evolutionary time, which are obtained
from the simulation data
using a slightly modified version of the {\tt GalevNB} program
\citep{pang2016}. The BSS formation
rate from stellar collisions is much higher after
the core collapse (\cf Fig.~\ref{fig:15K_single_evol}) owing to the high
central density and the rapid mass segregation
(at least locally, close to the cluster's center) of the most
massive (MS) stars that occurs nearly
simultaneously to the collapse. The mass segregation
is important here without which there would not have been
the marked increase of BSS formation right after the
collapse. Some BSSs as well get ejected from the cluster
due to dynamical interactions; this is likely
as they are the most massive members of the cluster
and therefore are most likely to participate in
close encounters.  

\begin{figure}
\includegraphics[width=6.5cm, clip=true]{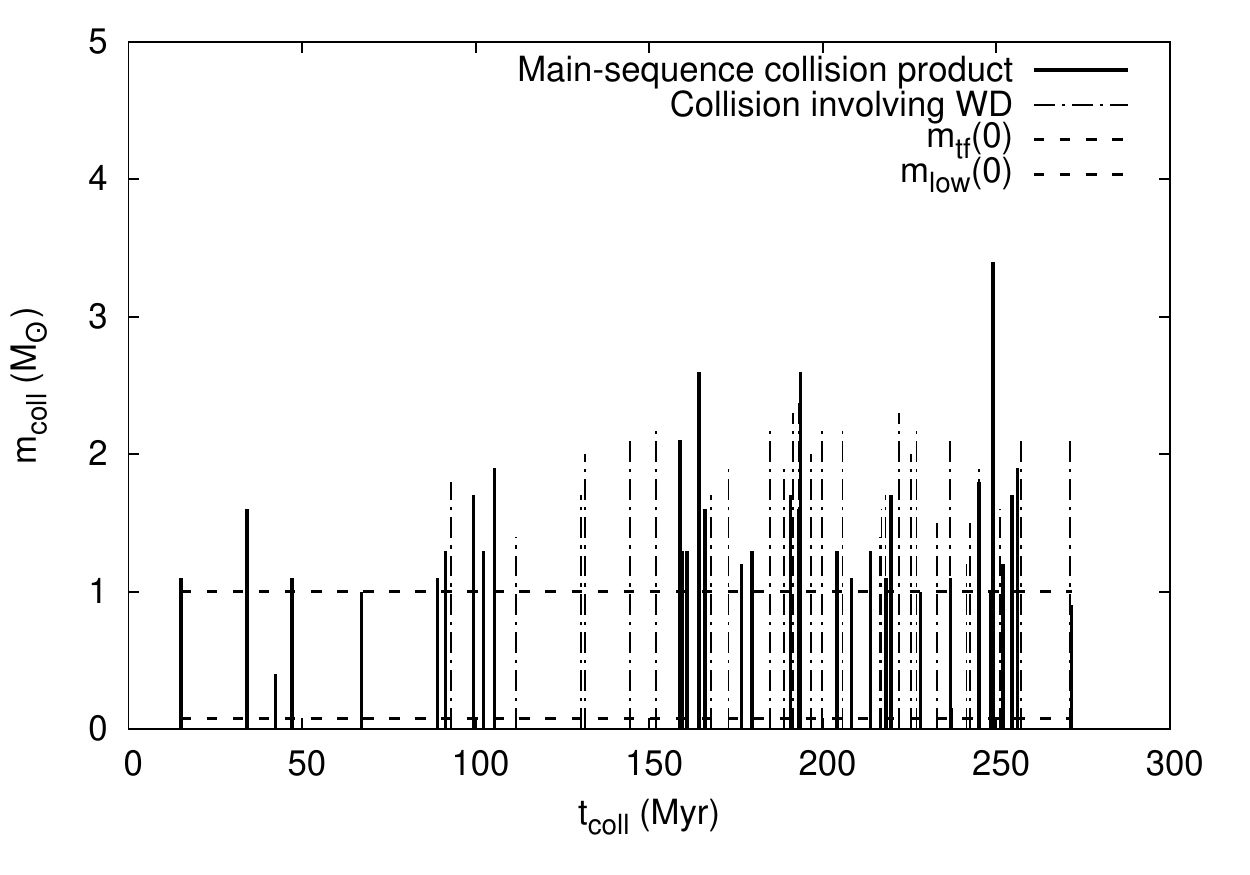}
\vspace{-0.9 cm}
\caption{\footnotesize Collision timeline for the calculation with
$\mcl(0)\approx3.0\times10^4\Ms$, $W_0=7.5$ and $\rh(0)=1$ pc
with initially only single stars.
}
\label{fig:30K_single_evol}
\end{figure}

The above condition is to some extent fine tuned as the
most massive member ($\approx 1.0\Ms$) is chosen to be
at the MS turn-off point
by adjusting the stellar pre-evolution
age. This has also made the cluster initially free of white dwarfs
(WDs) which can be more than or similarly massive as the turn-off mass
and might suppress the collision rate of MS stars close
to the turn-off, that generates the BSSs. To examine
the role of the WDs, a $\mcl(0)\approx3.0\times10^4\Ms$
cluster (no initial binaries) is computed with $0.08-8.0\Ms$ stars which
are pre-evolved for $\approx10$ Gyr; that gives
$\approx1\Ms$ turn-off mass and a large number of WDs. Although
the collisions between between WDs and MS stars
(resulting in red giants and AGBs) are frequent in this calculation,
BSSs continue to form and its rate boosts after the
core collapse. This is demonstrated in Fig.~\ref{fig:30K_single_evol}.

\begin{figure}
\flushleft
\includegraphics[width=5.2cm, angle=-90, clip=true]{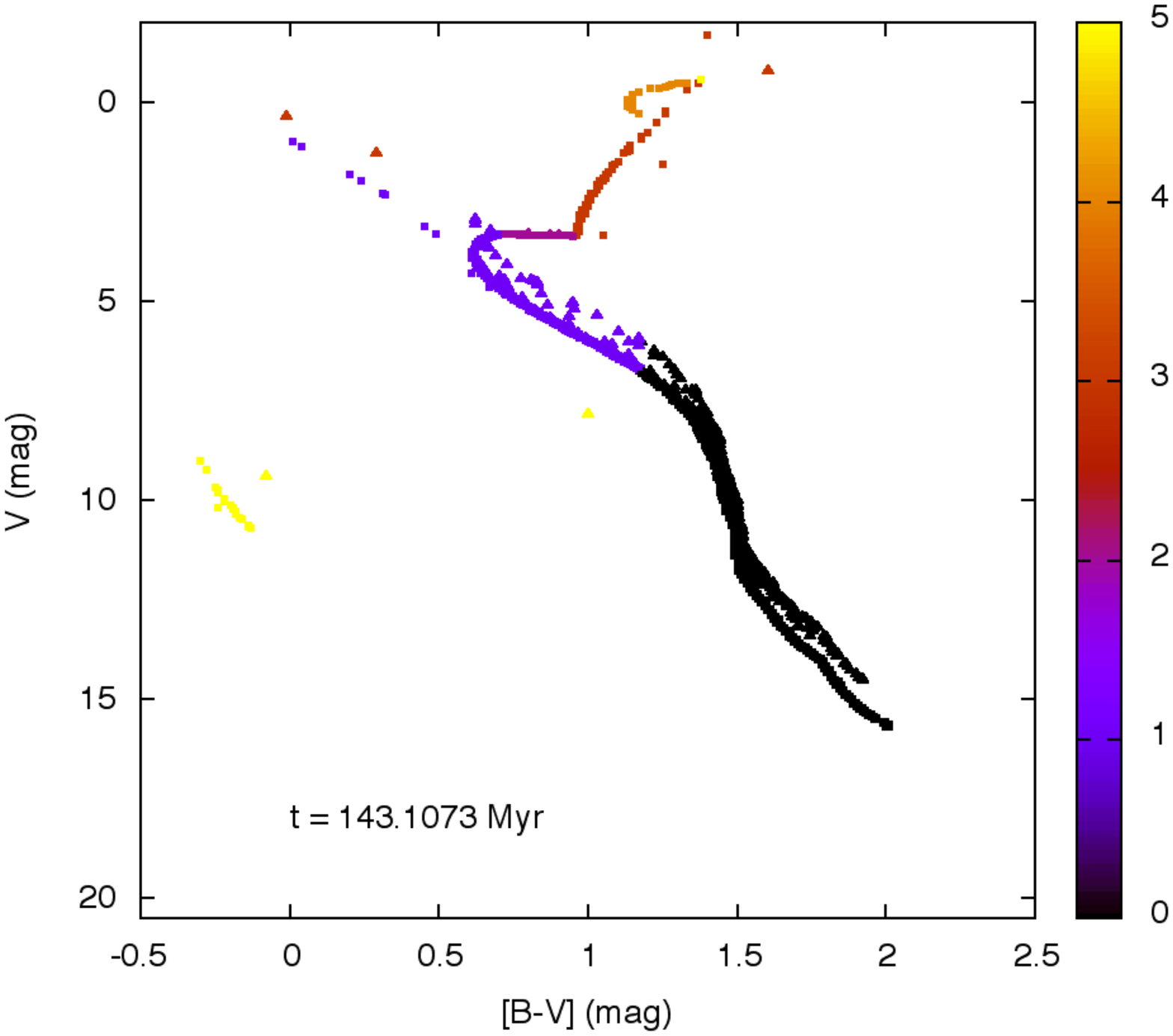}
\vspace{-0.4 cm}
\caption{\footnotesize The CMD at $t\approx143$ Myr evolution
for the calculation with
$\mcl(0)\approx3.0\times10^4\Ms$, $W_0=7.5$ and $\rh(0)=1$ pc
with initially $\approx5$\% binaries.
The meanings of the symbols and the colours are same
as in Fig.~\ref{fig:15K_single_cmd}.
}
\label{fig:15K_bin_cmd}
\end{figure}

In one final demonstration, a $\mcl(0)\approx1.5\times10^4\Ms$
cluster with $\approx5$\% binaries (see Sec.~\ref{calc}) is computed.
The corresponding CMD after $t\approx143$ Myr evolution is
shown in Fig.~\ref{fig:15K_bin_cmd}. A hint of double BSS sequence is apparent although
the number of BSSs in the red sequence is much less than that in
the blue sequence. On closer inspection of the computation,
it is found that both the binaries in the red sequence contain
a BSS and a red giant in a close binary having residual eccentricity.
This indicates a recent dynamical origin of these involving
encounters with the giant's envelope. This is consistent with
what has been envisaged so far. However, given that
only two of such binaries plus a third
MS-MS binary (close to the MS turn-off; see Fig.~\ref{fig:15K_bin_cmd})
define the red BSS sequence in this computed model, this is only a marginal
case of formation of a double BSS sequence.
This double BSS sequence is found to last for $\approx80$ Myr,
after which the binary BSSs evolve off.  
Computations of more massive clusters would potentially
provide more concrete outcomes.

The radial distribution of the BSSs in the model with primordial
binaries is also inspected where the members of the red
BSS sequence is found to be radially more concentrated then the
blue ones, as observed. Again, given that there are only three
red members, this comparison is only marginal. Interestingly,
in the computations without any initial binaries, the BSSs are much
more centrally concentrated than their counterparts in the model
with binaries --- the super-elastic encounters involving binaries
\citep{hh2003} and the resulting kicks make the BSSs more
radially spread out in the latter case.  

\section{Conclusions and outlook}\label{conclude}
\vspace{-0.32 cm}

From the above preliminary study suggest that:

\begin{itemize}

\item A ``burst'' of BSSs appears when an old (GC-like) cluster approaches
core collapse. This is true despite the presence of primordial binaries
(of a few percent) and a significant population of white dwarfs,
as long as ``some form of'' core collapse happens.

\item BSSs continue to form (and evolve/get ejected)
during post core collapse phase.

\item A ``second'' binary BSS sequence can appear in
the presence of primordial binaries which are typically
outcomes of recent dynamical interactions. 

\item Primordial binaries also
seem to determine the radial distribution of BSS.

\end{itemize}

Computing more massive models with primordial binaries is necessary
to consolidate the above results and to
better understand the properties of post core collapse
BSSs, which will be the forthcoming step of this study.

\end{document}